\documentclass[12pt]{article}
\usepackage[dvips]{graphicx}
\def\thefootnote{\fnsymbol{footnote}}
\def\be{\begin{equation}}
\def\ee{\end{equation}}
\def\ba{\begin{eqnarray}}
\def\ea{\end{eqnarray}}

\newcommand{\nn}{\nonumber \\ }

\begin{document}
\begin{titlepage}
\thispagestyle{empty}
\vskip0.5cm
\begin{flushright}
MS--TPI--99--11
\end{flushright}
\vskip0.8cm

\begin{center}
{\Large {\bf Width of Rough Interfaces}}
\end{center}
\begin{center}
{\Large {\bf on Asymmetric Lattices}}
\end{center}

\vskip0.8cm
\begin{center}
{\large Klaus Pinn}\\
\vskip5mm
{Institut f\"ur Theoretische Physik I } \\
{Universit\"at M\"unster }\\ {Wilhelm--Klemm--Str.~9 }\\
{D--48149 M\"unster, Germany \\[5mm]
 e--mail: pinn@uni--muenster.de
 }
\end{center}
\vskip3.0cm
\begin{abstract}
\par\noindent
I present a calculation of the interfacial width 
within the capillary wave (Gaussian) approximation. 
The calculation 
is done on rectangular lattices of size $L_1$ times $L_2$, 
with periodic boundary conditions.

\end{abstract}
\end{titlepage}

\setcounter{footnote}{0}
\def\thefootnote{\arabic{footnote}}

\section{Introduction}

The two-dimensional interface separating the phases of a 3D binary
system can undergo a roughening transition \cite{Weeks}.  Below the
roughening temperature, the interface has a finite width in the
infinite volume limit. The rough phase, however, is characterized by
strong fluctuations of the interface position variables, and the
interface width diverges when the area of the surface goes to
infinity. Properties of the roughening transition have been
investigated in a variety of studies, see e.g.\ \cite{matching} and
references therein.  It is generally believed, and has been confirmed
in a number of numerical studies \cite{numer}, that the infrared (large
distance) properties of a rough interface can be described by
massless Gaussian modes. This concept was introduced by Buff et
al.~\cite{Buff} and is usually referred to as the capillary wave model
(CWM).  The CWM allows to make predictions, e.g., about the finite
size effects of various interface properties 
\cite{Gelfand,Privman}. Quite some attention has been devoted 
to asymmetric interfaces, where the two extensions $L_1$ and 
$L_2$ of a rectangular interface do not necessarily coincide 
\cite{shape,DietzFilk,Potts,2loop}. In 
\cite{DietzFilk,Potts,2loop}
an extended 
(non-Gaussian) CWM was employed with an action proportional 
to the interface surface. The interface shape dependence 
was investigated in a field theoretic setting in 
\cite{FieldTheory}. 

In this article, I present a calculation of the interface 
width on asymmetric rectangular lattices, within the 
CWM. Related calculations with a continuum 
cutoff were done in a gauge theory setting in \cite{FluxTubes}. 
Interestingly, the lattice result obtained in the present paper 
points the way to a significant simplification of the 
continuum result presented in \cite{FluxTubes}.

\section{The Lattice Capillary Wave Model}

Let the interface (without overhangs) be described by a 
``height'' function $\varphi_x$, where $x$ is a site 
of a two-dimensional grid with side lengths $L_1$ and 
$L_2$. We choose units such that the lattice spacing
is dimensionless and assume that it is one for both 
lattice directions. We employ periodic boundary conditions. 

Within the capillary wave approximation for the fluctuations of the 
height variables, expectation values are defined by  
$$
\langle {\cal O} \rangle 
= \lim_{m \rightarrow 0}
\frac{\int \prod_x d\varphi_x \, e^{-H_0(\varphi)} \, {\cal O}(\varphi)}
{\int \prod_x d\varphi_x \, e^{-H_0(\varphi)}} \, , 
$$
with the Hamiltonian (energy functional) 
$$
H_0(\varphi) = \frac{1}{2\beta} \sum_{\langle x,y\rangle}
\left( \varphi_x - \varphi_y\right)^2 
+ \frac{m^2}{2} \sum_x \varphi_x^2 \, . 
$$
The first term is a sum over all nearest neighbour bonds in 
the lattice. We define the interface width (or surface thickness) 
by 
$$
W^2 = \frac{1}{L_1 L_2} \sum_x 
\langle \left( \varphi_x - \varphi_{x_0}\right)^2 \rangle \, , 
$$
where $x_0$ is an arbitrary site of the lattice. 
The interfacial width can also be written as 
$$
W^2 = 2 \, \langle (\varphi_{x_0} - \phi)^2 \rangle \, , 
$$
where 
$$
\phi = \sum_x \varphi_x 
$$
is the average interface position. 
In the following section we will show that 
\be
\frac{W^2}{2\beta} = 
K + \frac{\ln L_1 L_2}{4\pi} - \frac{1}{\pi} 
\ln\left( u^{1/4} \, \eta(iu) \right) 
+ \frac{1}{L_1 L_2} Z(u) + O\left( (L_1 L_2)^{-2} \right) \, . 
\ee
$\eta$ is the Dedekind function. 
We will compute explicit expressions for the constant $K$ 
and the function $Z(u)$ that depends on the asymmetry-parameter 
$$
u \equiv \frac{L_2}{L_1} \, . 
$$

\section{Calculation of $W^2$}

By means of discrete Fourier transformation $W^2$ can easily shown 
to be 
$$
W^2 = \frac{2\beta}{L_1 L_2} \, \sum_{p \neq 0} 
\frac{1}{\hat p_1^2 + \hat p_2^2} \, . 
$$
The lattice momenta are defined by $p_i= \frac{2\pi}{L_i} \, n_i$, 
with $n_i = 0,\dots,L_i-1$, and 
$\hat p_i^2 = 4 \, \sin^2 \frac{p_i}{2}$. Let us rewrite 
the sum excluding $p=(p_1,p_2)=0$ in the following way: 
$$
\sum^{\prime} f(n_1,n_2) = \sum_{n_2=1}^{L_2-1} f(0,n_2) 
+ \sum_{n_1=1}^{L_1-1} \sum_{n_2=0}^{L_2-1} f(n_1,n_2) \, .  
$$
Employing the identity 
$$
\frac{1}{\sinh^2(x/2) + \sin^2(\omega/2)}
= \frac{2}{\sinh x} \sum_{n=-\infty}^{\infty}
e^{-x |n| - i \omega n} \, , 
$$
we can perform a first summation: 
\be 
\label{ttt}
\sum_{n_2=0}^{L_2-1}
\frac{1}{\sin^2\left(\frac{\pi n_1}{L_1}\right) 
+ \sin^2\left( \frac{\pi n_2}{L_2} \right)}
= \frac{2 L_2}{\sinh X_{n_1}}
\left( 1 + 2 \frac{e^{-L_2 X_{n_1}}}{1 - e^{-L_2 X_{n_1}}} \right) \, . 
\ee 
The quantity $X_{n_1}$ is defined by 
$$
\sinh\left( \frac{X_{n_1}}{2} \right) = 
\sin\left( \frac{\pi n_1}{L_1} \right) \, .  
$$
Performing a suitable limit, one can use 
eq.~(\ref{ttt}) to show that 
$$
\sum_{n_2=1}^{L_2-1}
\frac{1}{\sin^2\left(\frac{\pi n_2}{L_2}\right)} = \frac13 
\left(L_2^2-1 \right)  \, . 
$$
Putting things together, we find 
$$
\frac{W^2}{2\beta}
= 
\frac{1}{12} \frac{L_2^2-1}{L_1L_2}
+ 
\frac{1}{2L_1} \sum_{n_1=1}^{L_1-1} 
\frac{1}{\sinh X_{n_1}}
\left( 1 + 2 \frac{e^{-L_2 X_{n_1}}}{1 - e^{-L_2 X_{n_1}}} \right) \, . 
$$
We shall first study the sum 
$$
S = \frac{1}{2L_1} \sum_{n_1=1}^{L_1-1} \frac{1}{\sinh X_{n_1}} \, . 
$$
Note that 
$$
\sinh X_{n_1} 
= 2 \sinh\left(\frac{X_{n_1}}{2} \right) 
\cosh\left(\frac{X_{n_1}}{2} \right) 
= 2 
\sin\left( \frac{\pi n_1}{L_1}   \right)
\sqrt{1 +\sin^2 \left( \frac{\pi n_1}{L_1}\right) } \, .
$$
Exploiting that 
$\sin\left( {\pi n_1}/{L_1}   \right)
= \sin\left( {\pi (L_1-n_1)}/{L_1}   \right)$
and assuming that $L_1 \equiv 2 M_1$ is even, we obtain 
\be
\label{abc} 
S  = 2 \, \sum_{n_1=1}^{M_1 - 1} g(X_{n_1}) + g(M_1)  \, , 
\ee
with $g(x_{n_1})=(2L_1 \sinh(X_{n_1}))^{-1}$.
$S$ can then be represented as $S = S_1 + S_2$, where 
$$
S_1  = 
\sum_{n_1=1}^{M_1-1}
\left( 
\frac{1}
{2L_1 
\sin\left( \frac{\pi n_1}{L_1}   \right)
\sqrt{1 +\sin^2 \left( \frac{\pi n_1}{L_1}\right) } }
-\frac{1}{2\pi n_1}
\right) + \frac{1}{4\sqrt{2} L_1}  \, , 
$$
and 
$$
S_2 = \frac{1}{2 \pi} \sum_{n_1=1}^{M_1-1} \frac{1}{n_1} \, . 
$$
The sum $S_1$ can be evaluated with the help of the Euler-Mc-Laurin 
summation formula. One finds 
$$
S_1 = \frac{1}{2\pi}
\left( \frac32 \ln 2 - \ln \pi 
+ \frac{1}{L_1} + 
\left( \frac13 + \frac{\pi^2}{36} \right) \,  \frac{1}{L_1^2}
\right) 
+ O\left(L_1^{-4} \right) \, . 
$$
$S_2$ can also be summed: 
$$
S_2 = \frac{1}{2\pi} \left( C + \psi(M_1) \right) \, , 
$$
where $C= 0.5772...$ denotes Euler's constant, and $\psi$ is the 
Digamma function. We expand 
$$
S_2 = \frac{1}{2\pi} \left(-\ln 2 + \ln L_1 + C - \frac1L_1 - \frac{1}{3L_1^2} 
\right) 
+ O\left(L_1^{-4} \right) \, . 
$$
$S$ thus combines to 
$$
S= \frac{1}{2\pi} \left( 
C + \frac12 \ln 2 - \ln \pi  + \ln L_1 \right) + \frac{\pi}{72 L_1^2}
+ O\left(L_1^{-4} \right) \, . 
$$
Our next task is to study 
$$
T = 
\frac{1}{L_1} \sum_{n_1=1}^{L_1-1} 
\frac{1}{\sinh(X_{n_1})}
\frac{e^{-L_2 X_{n_1}}}{1 - e^{-L_2 X_{n_1}}}  \, . 
$$
As for eq.~(\ref{abc}), we use the symmetry of the ``integrand'' 
to write 
\be
\label{qrs}
T = 
\sum_{n_1=1}^{M_1-1} h(X_{n_1}) 
+ h(X_{M_1}) \, , 
\ee
with 
$$
h(X_{n_1}) = 
\frac{2}{L_1}
\frac{1}{\sinh(X_{n_1})}
\frac{e^{-L_2 X_{n_1}}}{1 - e^{-L_2 X_{n_1}}}  \, . 
$$
Recall that $M_1=L_1/2$, and that we have assumed that $L_1$ is even. 
We are interested in the limit $L_1 \rightarrow \infty$, with 
the ratio $u = L_2/L_1$  kept fixed. 
One first observes that the contribution $h(X_{M_1})$ vanishes 
exponentially fast in this limit, namely like 
$\exp(-2 \, {\rm arsinh}(1)\, u L_1)$.  
For large $L_1$, we expand 
$$
h(X_{n_1}) = \frac1\pi \, 
\frac{q^{n_1}}{n_1 \, (1-q^{n_1})} 
\, \left( 1 + A(n_1,L_1,u) \right) \, , 
$$
where 
$$
q \equiv \exp(-2\pi u) \, . 
$$
We obtain 
$$
A(n_1,L_1,u)   
=
\frac{1}{L_1^2} \, \frac{\pi^2 n_1^2}{3} \, 
\left( 
\frac{2 \pi n_1 u}{1-q^{n_1}} - 1 
\right) + 
O\left( L_1^{-4} \right) 
 \, . 
$$
Extending the sum in eq.~(\ref{qrs}) to the range 1 to infinity 
introduces only errors that are exponentially small in $L_1$. 
We thus find 
$$
T = \frac1\pi 
\sum_{n=1}^{\infty} \frac{q^n}{n \, (1-q^n)} 
+ 
\frac{1}{L_1^2} \left( 
\frac{2 \pi^2}{3} \, u  \sum_{n=1}^{\infty}
\frac{n^2 \, q^n}{(1-q^n)^2} 
-
\frac{\pi}{3} \, \sum_{n=1}^{\infty}
\frac{n \, q^n}{1-q^n} 
\right) 
+ O\left( L_1^{-4} \right) \, . 
$$
Let us define 
$$
\begin{array}{ll}
F(u)& \hspace{-3mm}=  \sum_{n=1}^{\infty} \frac{q^n}{n \, (1-q^n)} \, , \\[5mm]
G(u)& \hspace{-3mm}= \sum_{n=1}^{\infty}
\frac{n^2 \, q^n}{(1-q^n)^2} \, , \\[5mm]
H(u)& \hspace{-3mm}= 
\sum_{n=1}^{\infty}
\frac{n \, q^n}{1-q^n} \, .
\end{array}
$$
Noting that $\ln L_1 = \frac12 \ln(L_1L_2) - \frac12 \ln u$, and 
putting everything together, we obtain 
\be
\label{theform} 
\frac{W^2}{2\beta} = 
K + \frac{\ln L_1 L_2}{4\pi} - \frac{1}{\pi} 
\ln\left( u^{1/4} \, \eta(iu) \right) 
+ \frac{1}{L_1 L_2} Z(u) + O\left( (L_1 L_2)^{-2} \right) \, . 
\ee
The constant $K$ is given 
by 
$$
K= \frac{1}{2\pi} 
\left( C + \frac12 \ln 2 - \ln\pi \right) \, . 
$$
Furthermore, we have used that $F(u)= G_1(q)$, where 
$$
G_p(x)= \sum_{n=1}^{\infty}
\frac{x^n}{n^p \, (1-x^n)}  
$$
is Lambert's series \cite{Apostol}. For $p=1$, it is related to Dedekind's 
$\eta$-function, 
$$
G_1\left( e^{-2\pi u} \right) 
= - \frac{\pi u}{12} - \ln \eta(iu) \, , 
$$
with 
$$
\eta(iu) = q^{1/24} \, \prod_{n=1}^{\infty} (1-q^n) \, .
$$
The symmetry of the interface width under exchange of 
$L_1$ and $L_2$ should be reflected in an invariance 
under the transformation $u \rightarrow 1/u$. 
It is well known that $u^{1/4} \eta(iu)$ is invariant under 
this transformation. 

The part proportional to $(L_1L_2)^{-1}$ is given by 
$$
Z(u)= \frac{\pi}{72} u - \frac1{12} 
+ \frac{2\pi^2}{3} u^2 \, G(u) - \frac{\pi}{3} \, u H(u) \, . 
$$
We would like to demonstrate that this expression is invariant 
under $u \rightarrow 1/u$. To this end we first observe 
that 
$$
H(u)= \frac{1}{24} \left( 1 - E_2(iu) \right) \, , 
$$
with $E_2$ being the first Eisenstein series. 
It obeys the functional relation 
\be
\label{trans}  
E_2(iu)= - u^{-2} E_2(iu^{-1}) + \frac{6}{\pi} \, u^{-1} \, . 
\ee 
The crucial step is now to recognize that $G$ can be written as 
$$
G = q\frac{d}{dq} H = - \frac{1}{48 \pi} \frac{d}{du} E_2(iu) \, . 
$$
We then have 
\be
\label{zz}
Z(u) = \frac{\pi}{72} \left( 
u E_2(iu) + u^2 \frac{d}{du} E_2(iu) \right) - \frac1{12} \, . 
\ee 
Differentiating 
eq.~(\ref{trans})
with respect to $u$ yields  
the behaviour of $\frac{d}{du} E_2(iu)$ under $u \rightarrow 1/u$.  
Using this and eq.~(\ref{trans}), it is easy to demonstrate 
the invariance of the right hand side of eq.~(\ref{zz}). 

We remark that it would be easy (though technical) to extend
eq.~(\ref{theform}) to higher orders in $(L_1 \, L_2)^{-1}$.

A comparison with eq.~(A.4) of ref.~\cite{FluxTubes} leads 
to an interesting observation. For the interface thickness 
on the continuum torus, 
regularized by a point splitting procedure, the authors obtain 
(I adapted their notation to the present setting) 
\begin{eqnarray}
\label{torino}
2\pi \sigma W^2=&\ln\left({L_1\sqrt{1+u^2}}/{2\varepsilon}
\right)+\frac1{2u}\arctan u + \frac u2\arctan\frac1u-\frac32
-\frac{\pi u}{12}\nn
-&\sum_{k=1}^{\infty}\frac{(1+u^2)E_{2k}(iu)}
{2ku(2k+2)!}\left((1+u^2)\pi^2\right)^kB_k'\sin\left((2k+2)\arctan 
u\right) \, .
\end{eqnarray}
Here, 
$$
E_{2k}(iu)= 1 + (-1)^k \frac{4k}{B_k'} \sum_{n=1}^{\infty}
\frac{n^{2k-1}q^n}{1-q^n} 
$$
is the $k$'th Eisenstein series, and $B_k'$ are Bernoulli 
numbers, defined through 
\be
\label{bernoulli}
\frac{e^z}{e^z-1} = 1 - \frac{z}2 
- \sum_{k=1}^{\infty} (-1)^k \frac{B_k'}{(2k)!} z^{2k} \, .  
\ee

Comparing eq.~(\ref{torino}) 
with the infinite area limit of eq.~(\ref{theform}), 
and identifying $\sigma \equiv 1/(2\beta)$, one is led 
to conjecture the identity 
\be 
\label{claim}
\frac12\ln(1+u^2) + g(u) = - 2\ln\eta(iu) - C \, , 
\ee 
with 
\begin{eqnarray}
g(u)= &\frac1{2u}\arctan u + \frac u2\arctan\frac1u-\frac32
-\frac{\pi u}{12}\nn
-&\sum_{k=1}^{\infty}\frac{(1+u^2)E_{2k}(iu)}
{2ku(2k+2)!}\left((1+u^2)\pi^2\right)^kB_k'\sin\left((2k+2)\arctan 
u\right) \nn \, . 
\end{eqnarray}
Eq.~(\ref{claim}) turns out to be true, with $C = -ln(\pi)$, thus 
leading to an enormous simplification of the result 
of \cite{FluxTubes}.
I will only give a sketch of the proof. 
One uses that 
$$
\sin((2k+2)\,   \arctan(u)) = 
(1+u^2)^{-k-1} {\rm Im}\left( 1 + iu)^{2k+2} \right) \, . 
$$
One then meets the following two expressions to be evaluated: 
$$
Y = 
\frac{1}{\pi^2 u} {\rm Im}
\sum_{k=1}^{\infty} 
\frac{B_k'}{2(2k+2)!} \left( \pi (1+iu) \right)^{2k+2}
$$
and 
$$
V = 
- \frac{2}{\pi^2 u} {\rm Im}
\sum_{n=1}^{\infty}
\frac{q^n}{n^3(1-q^n)}
\sum_{k=1}^{\infty} 
\frac{(-1)^k}{(2k+2)!} \left( n \pi (1+iu) \right)^{2k+2}
$$
The latter expression can be evaluated by first doing the $k$-sum 
(yielding a $\cos$ minus the two leading terms) and then 
extracting the imaginary part. One then again recognizes 
Lamberts series with $p=1$, and ends with 
$$
V = -\frac{\pi u}{6} - 2 \ln \eta(iu) \, . 
$$
For the study of $Y$, formula (A.6) of \cite{FluxTubes} is 
very helpful: 
$$
\sum_{k=1}^{\infty} (-1)^k 
\frac{B_k'}{2k(2k+2)!} z^{2k+2}
= \sum_{m=1}^{\infty} \frac{e^{-mz}-1}{m^3}
+ \frac{z^2}{2}\ln z + \frac{\pi^2}{6} z 
- \frac{3}{4} z^2 - \frac{z^3}{12} \, . 
$$
With $z= \pi(-u+i)$, and noting that for $u > 0$ 
$$
\ln z = \ln \pi + \frac12 \ln(1+u^2) - i\arctan(1/u) + i \pi \, , 
$$
one arrives after some algebra at 
$$
Y = -\ln\pi - \frac12\ln(1+u^2) +
\frac1{2u}\arctan\frac1u - \frac u2\arctan\frac1u+\frac32
+ \frac{\pi u}{4 }
- \frac{\pi}{4 u } \, . 
$$
Combining things and noting that (for $u > 0$) 
$\arctan u + \arctan \frac1u = \pi/2$, one arrives at eq.~(\ref{claim}). 

\section{Concluding Remarks}
The calculation presented in this paper is interesting 
at least for five reasons. First, it constitutes an illustrative 
example of analytical techniques that allow the exact evaluation 
of certain lattice sums. Second, the invariance under 
the exchange of the two lattice directions leads necessarily to the 
occurrence of modular forms and interesting relations between them. 
Third, the lattice result helped to simplify very much the 
continuum computation of ref.~\cite{FluxTubes}. Fourth, 
interesting physical applications are possible, e.g., in the context 
of analytical and Monte Carlo calculations 
(cf.\ again \cite{FluxTubes}). 
Finally, the present approach 
gives the the 1/area
correction to the asymptotic behaviour of the interface thickness. 
This could be useful as a 
starting point to test the CWM ``beyond the gaussian approximation'' 
\cite{Potts,2loop} also for the interfacial width.

\end{document}